\documentstyle[preprint,aps,prl]{revtex}

\begin{document}
\draft
\title{Observation of Strong Coulomb Blockade in Resistively Isolated Tunnel
Junctions}
\author{Wei Zheng, Jonathan R. Friedman, D.V. Averin, Siyuan Han\cite{han}, and J.E.
Lukens}
\address{Department of Physics and Astronomy, SUNY at Stony Brook,\\
Stony Brook, NY 11794}
\date{\today}
\maketitle

\begin{abstract}
We report measurements of the Coulomb-blockade current in resistively
isolated ($R_{Isol}\gg h/e^{2})$ tunnel junctions for the temperature range $%
60\,$mK $<T<230\,$mK where the charging energy $E_{c}$ is much greater than
the thermal energy. A zero-bias resistance $R_{0}$ of up to $10^{4}R_{T}$
(the tunnel resistance of the bare junction) is obtained. For $eV\ll E_{c},$
the $I$-$V$ curves for a given $R_{Isol}$ scale as a function of $V/T$, with 
$I\propto V^{\alpha (R_{Isol})}$ over a range of $V.$ The data agree well
with numerical calculations of the tunneling rate that include environmental
effects.
\end{abstract}


\label{intro}Coulomb blockade in small-capacitance tunnel junctions has been
intensively studied in the past decade\cite{averin,grabert}. In general,
Coulomb blockade is observed when the electric charge $Q$ is localized on a
small conducting island, i.e. when both thermal and quantum fluctuations of
the charge are suppressed. Thermal fluctuations are suppressed when the
temperature is below the charging energy $E_{c}=e^{2}/2C$ of the island
capacitance $C$. Suppression of quantum fluctuations requires the isolation
resistance $R_{Isol}$ between the island and the ``outside world'' to be
much greater than the quantum resistance $R_{K}=h/e^{2}$. In systems of
several tunnel junctions, where the electrodes are physically isolated by
tunnel barriers, this condition is easily satisfied, leading to strong
Coulomb blockade\cite{geerligs}. However, in a single, current-biased tunnel
junction, the localization of charge can only be achieved in practice with
an Ohmic resistor\cite{delsing}. This requires the fabrication of a compact
resistor of large magnitude, $R_{Isol}\gg R_{K}$. The major problem in
achieving this is that the isolating resistor acts as a lossy (RC)
transmission line (unless it is very short), with an impedance $%
Z_{Isol}(\omega )\ll $ $R_{K}$ for frequencies approaching $E_{c}/{\hbar }$,
which are important for the tunneling process. Previously published reports
on resistively isolated junctions\cite{clarke,kuzmin,holst,pekola,popovic,joyez}
showed relatively small increases in the zero-bias resistance $R_{0}$
compared to the high-bias tunnel resistance $R_{T}$. In this Letter we
report the observation of strong ($R_{0}/R_{T}$ up to $10^{4})$ Coulomb
blockade in a resistively isolated junction, and the first observations of
the predicted power-law variation of the tunneling current $I(V)$ in the
blockade region with an exponent dependent on $R_{Isol}$, in analogy to
recent observations of similar behavior in Luttinger liquids\cite
{kane,milliken,chang,tsui}. This strong blockade enables us to
quantitatively test the ``environment'' theory, based on the electrodynamic
description of the resistor.

The theory for the effect of such resistive environments (characterized by
continuous charge transfer) on the tunneling rate through the junction is
now well developed for the case $R_{T}\gg R_{K}$\cite
{nazarov,devoret,girvin,ingold}. These calculations predict that at zero
temperature and for $V\ll V_{c}$ ($V_{c}\equiv E_{c}/e)$ the current varies
as a power of energy change $E$ of the system when an electron tunnels.
Since, for a single junction, $E=eV$, this gives: 
\begin{equation}
I\propto E^{\alpha }\propto V^{\alpha },\hspace{0.5in}\alpha \equiv \frac{%
2R_{Isol}}{R_{K}}+1.  \label{powerI}
\end{equation}
For finite (but sufficiently low) temperatures, the zero-bias resistance $%
R_{0}$ varies as a power $T:$ 
\begin{equation}
R_{0}\propto T^{1-\alpha }.  \label{powerT}
\end{equation}
The origin of these power laws is the one-dimensional nature, i.e. the
linear variation with frequency, of the number of the ``photon modes'' used
to model the resistors. In this respect, they are analogous to the power law
($I\propto V^{\alpha _{LL}})$ recently predicted and observed for tunneling
in Luttinger liquids\cite{kane,milliken,chang,tsui}. The two systems are
very different on the microscopic level: a strictly 1D ballistic conductor
for the Luttinger liquid, and a disordered macroscopic conductor with a
large number of transverse electron modes leading to 3D electron motion for
the Ohmic resistor. Nonetheless, the correlation functions of both systems
are dominated by bosonic excitations with a linear spectrum. This particular
spectrum gives rise to the power-law $I$-$V$ characteristics.

\begin{figure}[tbh]
\epsfxsize=6.5in
\epsfysize=6.0in
\epsfbox{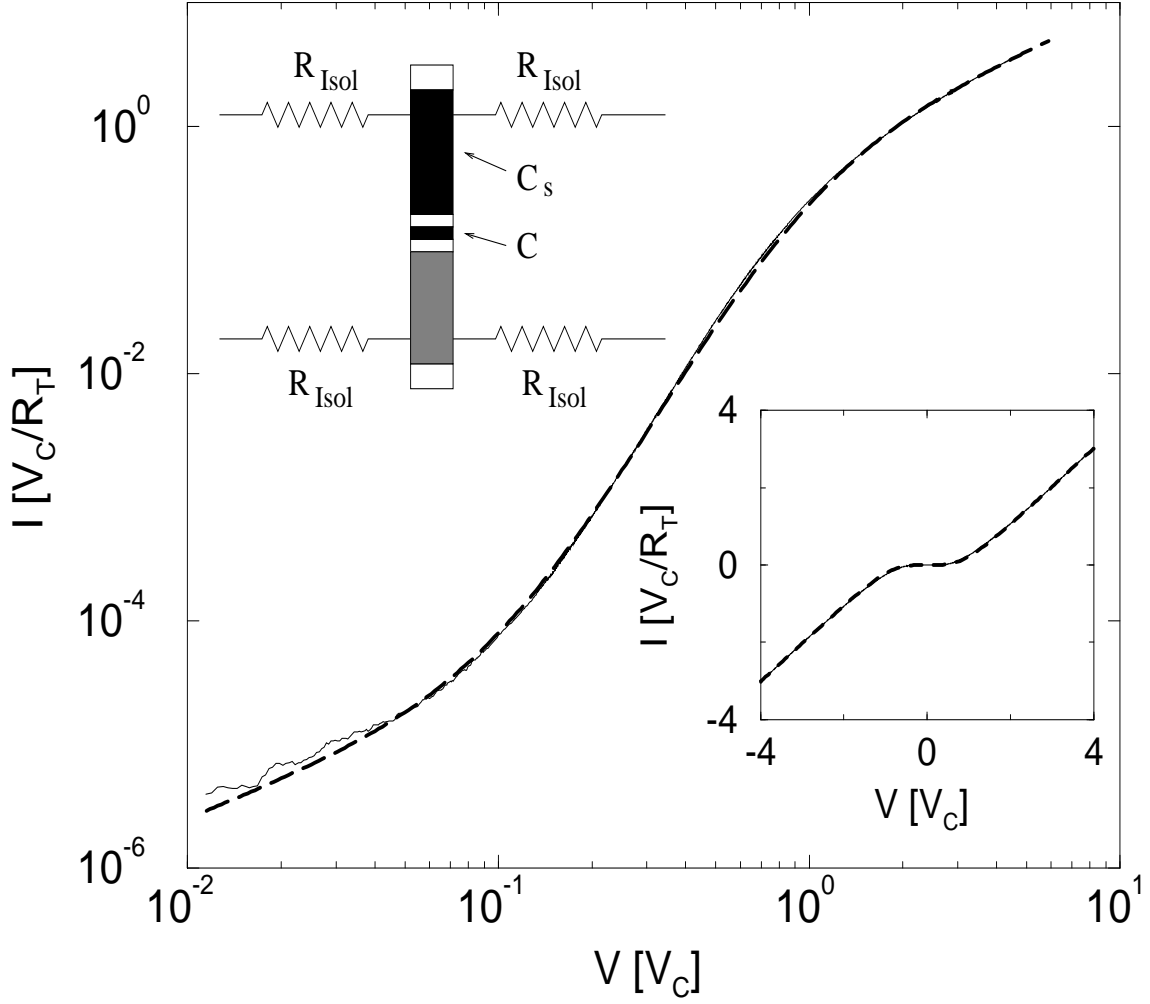}
\caption{$I$-$V$ characteristics (thin solid line) of sample 1 at 70mK
compared with the theoretical fit (thick dashed line) using $R_{Isol}$ as a
free parameter. The upper inset shows the schematic of the sample,
where the shaded areas are the overlapping regions formed by Al shadow
evaporation and the darkest areas indicate the junctions. $C$ and $C_s$
are the capacitances of the main junction and the secondary junction
respectfully, and $R_{Isol}$ represent the isolating resistors. The
lower inset gives the linearly scaled $I$ -$V$ curve and associated fit. }
\label{iv}
\end{figure}

The sample (see upper inset of Fig.\thinspace \ref{iv}) is fabricated with a
three-level process. First, the resistor level is patterned and a bilayer
film of 3nm Cr and 25nm of Au is sequentially deposited without breaking
vacuum. An etch window is then patterned over parts of the leads and Au
removed to form the Cr wires, which are approximately 100nm wide and have a
resistivity of about 10 k$\Omega /\mu $m. Finally, the junction-level mask
is patterned, and the standard shadow-evaporation technique with a vertical
offset is used to form the Al/AlO$_{x}$/Al tunnel junctions (dark rectangles
in upper inset). The specific capacitance of 4.5 fF/$\mu $m$^{2}$ gives a
nominal capacitance for the small junction of about $C=$ 230 aF. Adding to
this is the parasitic capacitance $C_{p}$ of the leads between the junction
and the resistors, which we estimate to be $\sim $35 aF based on
calculations of the self capacitance of other structures with similar wires 
\cite{kenji}. The spaces between the resistors are 1.5 $\mu $m horizontally
and 1.5 $\mu $m vertically. Calculations show that the isolating resistors
can be treated as ideal lumped resistors for frequencies up to 100\thinspace
GHz. The resistors have nearly temperature-independent, linear $I$-$V$
characteristics to the lowest temperature. Their noise properties, measured
down to 1.2 K, are found to be consistent with Johnson noise for the range
of bias currents used in the data presented below. Measurements are made
with the sample enclosed in a liquid-He-filled copper cell located on the
temperature-regulated platform of a dilution refrigerator. All leads
entering this cell pass though high-attenuation microwave filters\cite{dan},
which are thermally anchored to the platform. A magnetic field of
1\thinspace T is applied to suppress superconductivity in the sample.

The fabrication process, chosen to minimize $C_{p}$, results in a secondary
series junction with an area (and capacitance $C_{s}$) about 15 times that
of the small junction, making the device into an asymmetric transistor,
albeit one whose conductance is almost entirely determined by the small
junction. The energy change as an electron tunnels through the small
junction is given by $E(q)=\kappa \lbrack eV+(q-e/2)/C_{s}]$, where $q$ is
the island charge and $\kappa \equiv C_{s}/(C+C_{s})\simeq 1$. For the
region of interest, $V\ll V_{c}$, where the current is determined by the
forward tunneling rate through the small junction, the voltage for a fixed
bias current is minimized for $q=e/2$. In this case, $E(e/2)=\kappa eV$ and,
within the sequential tunneling model, the asymmetric transistor is
predicted to behave essentially as a single junction\cite{ingold}: It has no
blockade for $R_{Isol}\rightarrow 0$, has a blockade voltage $V_{c}=e/2C$
when $R_{Isol}\gg R_{K}$, and exhibits the power-law behavior of
Eqs.\thinspace \ref{powerI} and \ref{powerT} with $\alpha \rightarrow \kappa
^{2}\alpha $. Both the smaller ``bare'' resistance of the secondary junction
and the much weaker effect of the high impedance environment ($\kappa
_{s}^{2}<0.01$) ensure that the first junction is the bottleneck for the
electron transfer across the device. If the resistance $R_{s}$ of the
(probably leaky) secondary junction is much less than $h/e^{2}$, $\kappa
\rightarrow 1$, and the correspondence between the transistor with $q=e/2$
and the single junction is even closer, with $R_{s}$ simply increasing $%
R_{Isol}$ slightly.

\begin{figure}[htb]
\epsfxsize=6.5in
\epsfysize=6.0in
\epsfbox{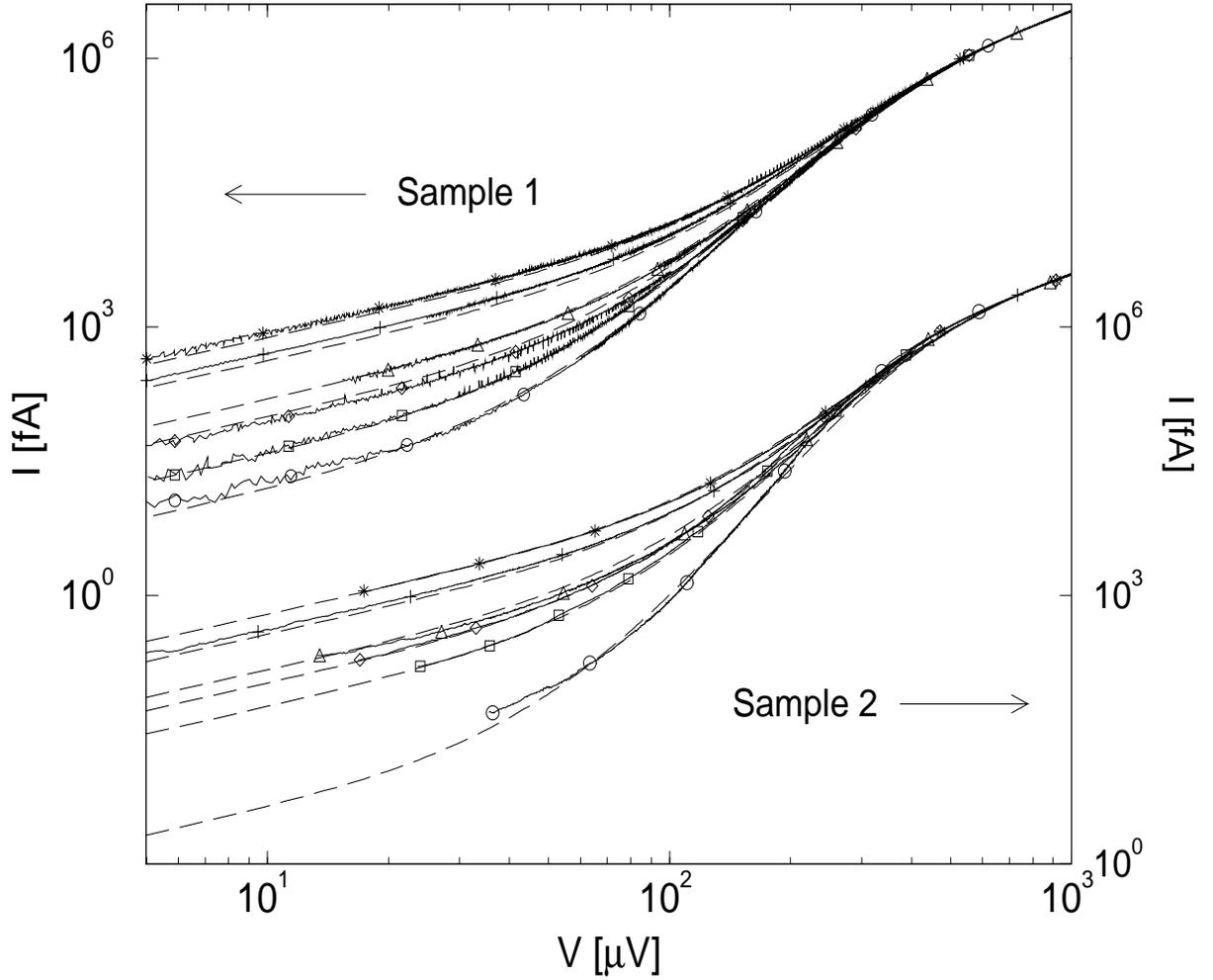}
\caption{$I$-$V$ curves at different temperatures for sample 1(2) compared
with numerical results from the full theory. The temperatures are 70, 90,
118, 135, 184 and 225mK (bottom to top) for sample 1 and 75, 90, 122, 143,
195and 225mK (bottom to top) for sample 2.}
\label{rj1}
\end{figure}

We present data from two samples with nominally identical junctions but with
isolating resistors of different length and resistance: 4 $\mu $m
(40\thinspace k$\Omega )$ for sample 1 and 8 $\mu $m (75 k$\Omega )$ for
sample 2. The parameters for these samples are summarized in Table \ref
{table1}. All data are taken with $q=e/2$ as determined by adjusting the
back-plane gate to minimize $V$ for $V\ll V_{c}$. A low-temperature $I$-$V$
curve for sample 1 is shown in Fig.\thinspace \ref{iv}. The voltage offset, $%
V_{c}=455\,$\thinspace $\mu $V, due to Coulomb blockade is clearly seen in
the linear-scale lower inset. The current variation for $V<V_{c}$ is more
clearly seen in the main log-log plot, where it is compared with the
numerical solution of the theory. The large current suppression ($%
R_{0}>4\times 10^{3}R_{T}$) indicates the strong blockade seen in this
sample. Figure\thinspace \ref{rj1} shows $I$-$V$ curves taken at several
temperatures for both samples, again compared with the result of numerical
solution. One can see that sample 2, which has a larger $R_{Isol}$, has a
stronger blockade and a steeper power-law region at the lowest temperature.

The current $I(V,T)$ (dashed lines in Figs.\thinspace \ref{iv} and \ref{rj1}%
) is calculated using the standard golden-rule approach with the effects of
the environment included through the correlation function of phase
fluctuations due to the total environmental impedance $Z_{t}(\omega )$\cite
{nazarov,devoret,girvin}. We take the real part of this impedance to be: Re$%
\left[ Z_{t}(\omega )\right] =R_{Isol}/(1+\left( {\omega }/{\omega }%
_{RC}\right) ^{2})$ where $\omega _{RC}=1/R_{Isol}C_{e}$ and $C_{e}$ is the
effective junction capacitance corresponding to the experimentally
determined value of $E_{c}$ (see Table \ref{table1}). $E_{c}$ and $R_{T}$
are determined from the high-voltage data, where the corrections due to
finite temperature and $R_{Isol}$ are negligible. The isolating resistance $%
R_{Isol}$ is then used as the only adjustable parameter in fitting the data
for $V<V_{c}$, giving an excellent fit over six orders of magnitude in the
current. In view of the essentially identical behavior of the transistor
with $q=e/2$ and the single junction, and the impossibility of accurately
determining $R_{s}$, we treat the sample as a single junction and admit some
uncertainty in our comparison of the measured and fitted values of $R_{Isol}$%
. Table \ref{table1} compares this best-fit value of $R_{Isol}$ with the
independently measured value. For both samples the best-fit value is
approximately 10 k$\Omega $ higher than the measured value. This may very
well be the resistance of $R_{s}$

As noted, the physical basis for the power-law behaviors seen in these
samples and in Luttinger liquid is the same. Therefore it is of considerable
interest to examine in more detail the region of validity for those
behaviors in our devices. While the full theory (Figs.\thinspace \ref{iv}
and \ref{rj1}) requires numerical integration, analytical expressions
illustrating the power-law behaviors in Eqs.\thinspace \ref{powerI} and \ref
{powerT} for the conductance $G$ can be derived for the limit $T\ll
T_{RC}\equiv {\hbar }{\omega }_{RC}/k_{B}$ and $V\ll V_{RC}\equiv {\hbar }{%
\omega }_{RC}/e$: 
\begin{equation}
G(V,T)\equiv \frac{I(V,T)}{V}=G_{0}(T)f\left( \frac{V}{T}\right) ,
\label{scale}
\end{equation}
where $G_{0}(T)$ is the zero-bias conductance given by 
\begin{equation}
G_{0}(T)=\frac{1}{R_{T}}\left( \frac{2\pi \text{k}_{B}T}{{\hbar }{\omega }%
_{c}}\right) ^{\hspace{-0.05in}\alpha -1}\frac{\left[ \Gamma \left[ \frac{1}{%
2}\left( \alpha +1\right) \right] \right] ^{2}}{\Gamma \left( \alpha
+1\right) }  \label{R0}
\end{equation}
with $\omega _{c}$(${\sim }\omega _{RC}$) being the cutoff frequency of
excitations, and 
\begin{equation}
f\left( \frac{V}{T}\right) =\left| \frac{\Gamma \left[ \frac{1}{2}\left(
\alpha +1\right) +i\frac{eV}{2\pi \text{k}_{B}T}\right] }{\Gamma \left[ 
\frac{1}{2}\left( \alpha +1\right) \right] \Gamma \left[ 1+i\frac{eV}{2\pi 
\text{k}_{B}T}\right] }\right| ^{2}.  \label{V/T}
\end{equation}
$\Gamma (x)$ is the Gamma function. Equation{\thinspace }\ref{R0} explicitly
shows the power-law temperature dependence of $G_{0}$($\equiv 1/R_{0}$) for $%
eV/2\pi $k$_{B}T\ll 1$. For $eV/2{\pi }k_{B}T\gg 1$, the Gamma functions in
Eq.{\thinspace }\ref{V/T} can be expanded, yielding the voltage power law $%
G(V)\propto V^{(\alpha -1)}$.

\begin{figure}[htb]
\epsfxsize=6.5in
\epsfysize=6.0in
\epsfbox{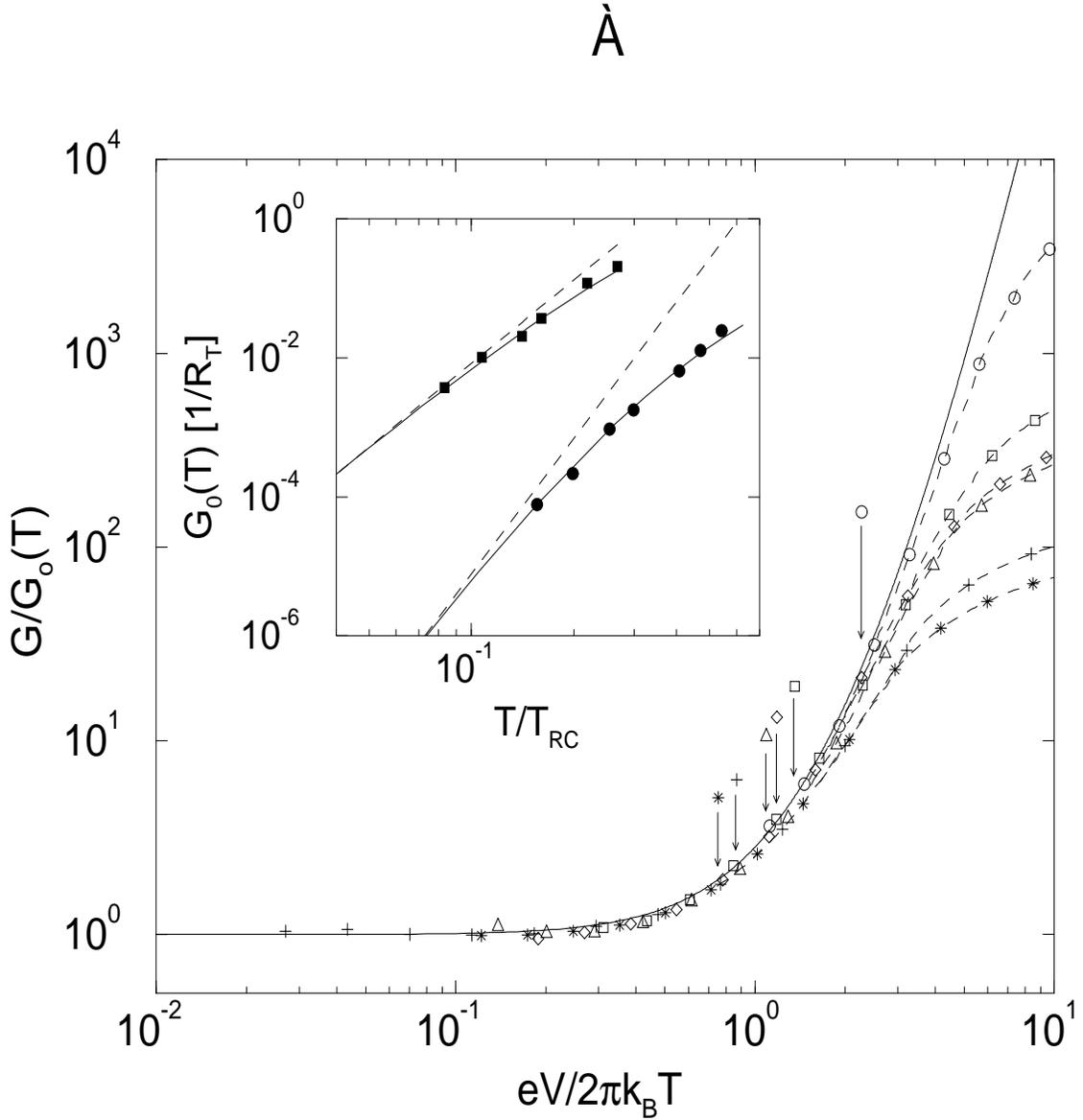}
\caption{Scaled conductance of sample 2 compared with the scaling function
given by Eq.\thinspace\ref{V/T} (solid line) for data at 75, 90, 122, 143,
195and 225mK (top to bottom). The arrows indicate the points where $V =
0.25V_c $ for the data set corresponding to the associated symbol. The inset
shows the dependence of $G_{0}$ on temperature for both samples (squares for
sample 1, circles for sample 2) compared with finite-temperature theory for
each sample (solid lines). The dashed lines, which are shifted to
asymptotically match the simulations, show the power-law behavior predicted
by Eq.\thinspace\ref{R0} for $T \ll T_{RC}$. }
\label{rj2}
\end{figure}

Equation\thinspace \ref{scale} shows that, at low bias and fixed $R_{Isol}$,
the conductance can be scaled to a function of $V/T$ as seen in Fig.{%
\thinspace }\ref{rj2}. The main figure shows this scaled conductance
compared to the prediction of Eq.\thinspace \ref{V/T} (solid line). The
arrows associated with the data sets at different temperatures indicate the
points at which $V=0.25V_{c}$, where deviations from the low-voltage
approximation begin to appear. For this scaling, $G_{0}(T)$ has been used as
a free parameter at each temperature. The resulting values for $G_{0}(T)$
for both samples are compared with the numerical results (solid lines) in
the inset of Fig.\thinspace \ref{rj2} (Note that $T$ is scaled to $T_{RC}$
for each sample.). Again, agreement is quite good. Sample 2, which has a
larger $R_{Isol}$ (see Table \ref{table1}), shows the predicted decrease in $%
G_{0}(T)$ and increase in the power-law exponent compared to sample 1.
However, the low-temperature approximation (Eq.\thinspace \ref{R0}, dashed
lines) deviates significantly from the data (and the finite-temperature
theory) at $T>0.1{\,}T_{RC}$. The scaling of voltage with temperature as
well as the power-law dependence of $I$ on $V$ are much more robust, with
deviations of less than 15\% from Eq.\thinspace \ref{V/T} for $V<0.25V_{c}$,
independent of temperature and $V_{RC}$. This fit, along with the strong
temperature dependence of $G_{0}(T)$, indicates that the heating in $%
R_{Isol} $ is not significant, at least for $V<0.25V_{c}$. For $T<60\,$mK,
the data become nearly independent of T. We cannot presently rule out
experimental artifacts as the explanation for this effect, which requires
further study.

In summary, we have measured the $I$-$V$characteristics of resistively
isolated tunnel junctions with $R_{Isol}\gg R_{K}$. We find strong Coulomb
blockade with an increase of up to 10$^{4}$ in the zero-bias resistance
compared to the high-bias tunnel resistance of the junctions. For
temperatures above 60{\thinspace }mK, the results agree well with the theory
based on an ``environmental'' description of the resistor as a collection of
oscillators with a linear spectrum. In particular we clearly observe the
predicted power-law dependences of $I(V)$ and $G_{0}(T)$ with the power-law
exponents dependent on $R_{Isol}$ in a manner consistent with the theory.

We thank K.K. Likharev and A. Rylyakov for useful and stimulating
discussions. This work was supported in part by AFOSR.

\input{epsf} 
\begin{figure}[tbh]
\end{figure}

\input{epsf} 

\input{epsf} 

\begin{table}[tbp]
\caption{Characteristics of the samples. }
\label{table1}
\begin{tabular}{cccc}
Sample No. & 1 & 2 &  \\ 
$R_T (k\Omega)$ & 170 & 152 &  \\ 
$E_c (K) $ & 5.28 & 4.89 &  \\ 
$V_c ({\mu}V) $ & 455 & 422 &  \\ 
$V_{RC} ({\mu}V) $ & 71 & 41 &  \\ 
$T_{RC} (K) $ & 0.84 & 0.48 &  \\ 
$R_{Isol} (meas.) (k\Omega)$ & 40 & 75 &  \\ 
$R_{Isol} (fit) (k\Omega)$ & 52 & 84 & 
\end{tabular}
\end{table}

\end{document}